\documentclass[a4paper]{article}

\usepackage[english]{babel}
\usepackage[utf8]{inputenc}
\usepackage[T1]{fontenc}

\usepackage{listings}
\usepackage{booktabs}

\usepackage{lmodern}
\usepackage[autostyle]{csquotes}

\usepackage[a4paper,top=3cm,bottom=2cm,left=3cm,right=3cm,marginparwidth=1.75cm]{geometry}

\usepackage{amsmath}
\usepackage{graphicx}
\usepackage[colorinlistoftodos]{todonotes}
\usepackage{authblk}
\usepackage{rotating}
\usepackage{csvsimple}
\usepackage{float}

\usepackage[colorlinks=true, allcolors=blue]{hyperref}
\usepackage{rotating}

\title{\huge A Critique of Differential Abundance Analysis, \\and Advocacy for an Alternative}%, and Why Algorithmic Multivariable Models Fair Better}
\author[1*]{Thomas P. Quinn}
\author[2]{Elliott Gordon-Rodriguez}
\author[3]{Ionas Erb}

\affil[1]{\footnotesize Applied Artificial Intelligence Institute (A2I2), Deakin University, Geelong, Australia}
\affil[2]{\footnotesize Department of Statistics, Columbia University, New York, USA}
\affil[2]{\footnotesize Centre for Genomic Regulation (CRG), The Barcelona Institute of Science and Technology, Barcelona, Spain

*\textit{contacttomquinn@gmail.com}
}
\date{}

\Affilfont{\fontsize{4}{4}}

\begin{document}
\maketitle

\begin{abstract}

It is largely taken for granted that differential abundance analysis is, by default, the best first step when analyzing genomic data. We argue that this is not necessarily the case. In this article, we identify key limitations that are intrinsic to differential abundance analysis: it is (a) dependent on unverifiable assumptions, (b) an unreliable construct, and (c) overly reductionist. We formulate an alternative framework called \textit{ratio-based biomarker analysis} which does not suffer from the identified limitations. Moreover, ratio-based biomarkers are highly flexible. Beyond replacing DAA, they can also be used for many other bespoke analyses, including dimension reduction and multi-omics data integration.

\end{abstract}

\maketitle

\section{What is differential abundance?}

\textit{Differential abundance analysis} (DAA), by which we mean the statistical analysis of single features, is commonly used to analyze high-throughput sequencing experiments. The intuition behind DAA is simple: A feature -- for example, the abundance of a gene -- may be relevant to an experimental condition if it associates with that condition. One can think of DAA in terms of a generalized linear model:
\begin{equation}
    \textbf{y} \approx \phi \Big( \beta_j \textbf{x}_j + \beta_0 \Big),
\end{equation}
where $\textbf{y}$ describes the experimental condition (e.g., case vs. control) and $\textbf{x}_j$ describes the abundance of gene $j\in \{1,\dots,G\}$ as measured across $i \in \{1,\dots,N\}$ samples.\footnote{In practice, most popular implementations of DAA will treat the abundance of a gene as a dependent variable and the experimental condition as the independent variable (corresponding to a retrospective study). We instead treat the experimental condition as the dependent variable (corresponding to a prospective study) in order to align DAA more clearly with the algorithmic modelling culture, for example to predict a disease status as a function of gene expression. Note that, under standard assumptions, inference in the prospective and retrospective models is typically equivalent \cite{prentice_logistic_1979}.} The optional transformation $\phi$ depends on the distribution of $\textbf{y}$. If $\textbf{y}$ is binary, $\phi$ may be a logistic transform. When appropriate, one could add covariates to adjust the model for potential confounders. A gene is said to have \textit{differential abundance} (DA) if its effect size $\beta_j$ exceeds some threshold, for example a cutoff based on a statistical test that gives an adjusted p-value less than 0.05.

\begin{figure}[H]
\centering
\scalebox{1}{
\includegraphics[width=(.8\textwidth)]{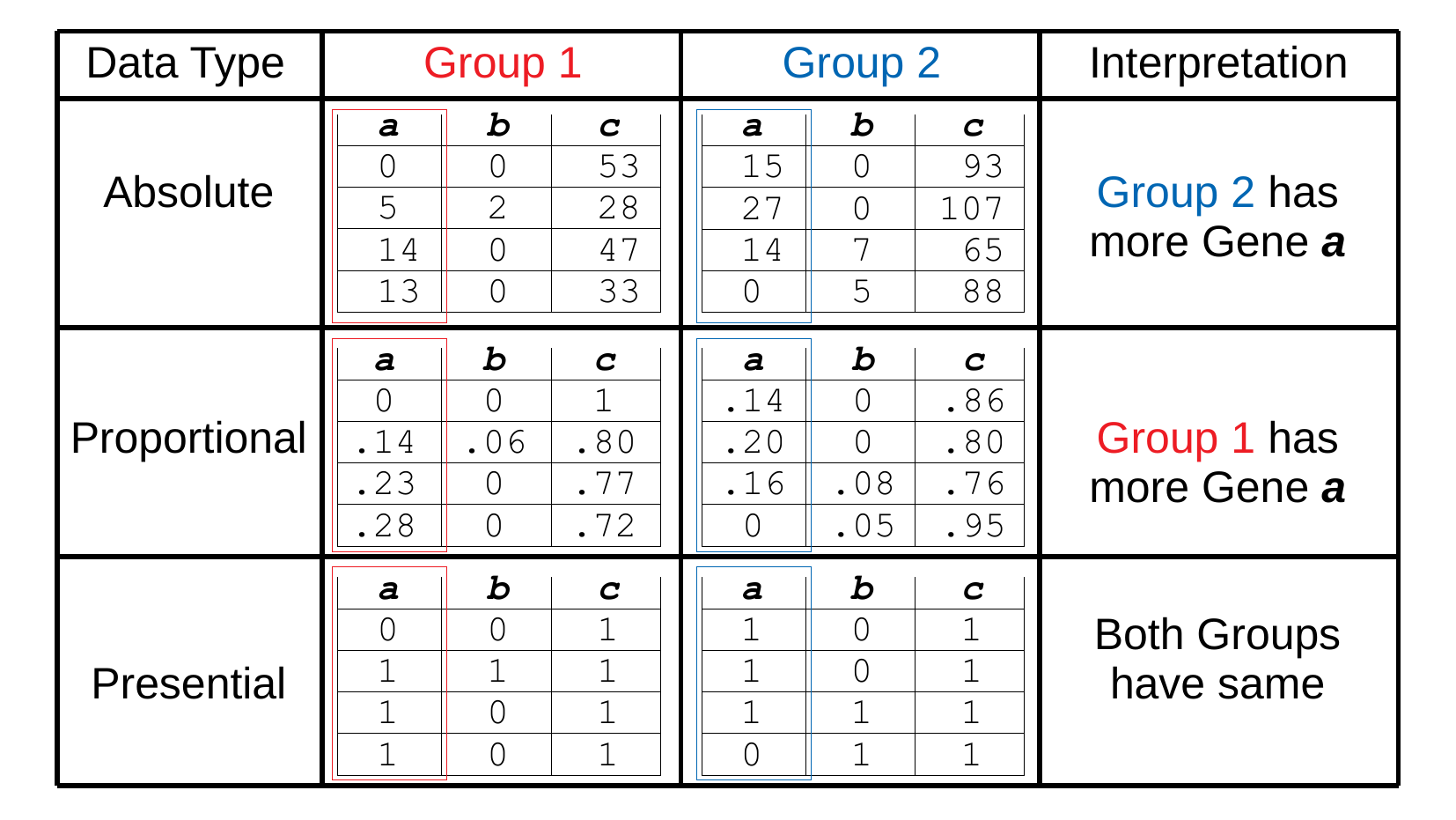}}
\caption{There are (at least) 3 different notions of differential abundance, depending on what the underlying data represent. These notions may not agree with one another. For example, a gene that is absolutely over-abundant in one group may appear otherwise when viewed in proportional or presential terms. Complicating matters further, sequencing biases typically make absolute abundances unavailable.}
\label{fig:datypes}
\end{figure}

In practice, the application of DAA is more nuanced. There exist (at least) 3 potentially conflicting notions of DA, depending on what $\textbf{x}_j$ actually describes. Differential abundance can be: \textit{absolute} when $\textbf{x}_j$ describes the true abundances of a gene, \textit{relative} when $\textbf{x}_j$ describes the proportional abundances of a gene, or \textit{presential} when $\textbf{x}_j$ describes the presence or absence of a gene. Figure~\ref{fig:datypes} shows how these notions of DA may disagree with one another. % or \textit{non-parametric} when $\textbf{x}_j$ describes the ranked abundances of a gene.

Many methods for DAA have been proposed, e.g, \cite{anders_differential_2010,robinson_edger:_2010,fernandes_anova-like_2013,paulson_differential_2013,law_voom:_2014,mandal_analysis_2015,kumar_analysis_2018}, (and keep being proposed, e.g., \cite{davis_rank_2021, zhou_linda_2021}) and several attempts have been made to systematically benchmark them against one another \cite{soneson_comparison_2013,dillies_comprehensive_2013,thorsen_large-scale_2016,hawinkel_broken_2017,weiss_normalization_2017,pereira_comparison_2018,quinn_benchmarking_2018}. However, DAA has received little scrutiny at the highest level of consideration. It is largely taken for granted that DAA is, by default, the best first step when analyzing genomic data. We argue that this is not necessarily the case. This article has 3 aims: (1) To summarize three popular approaches to DAA; (2) To highlight some limitations intrinsic to these approaches; and (3) to formulate an alternative framework called \textit{ratio-based biomarker analysis} that overcomes the limitations of DAA.

\section{Three Approaches to Differential Abundance}

There is a general consensus among bioinformaticians that DAA requires something more than a linear model: the abundances must first get \textit{pre-processed} to adjust for systematic biases. The pre-processing step is often called a \textit{normalization}. However, there are diverging opinions about how the data ought to be pre-processed \cite{mcmurdie_waste_2014} and what pre-processing steps ought to meet the definition of normalization \cite{quinn_understanding_2018}.

To understand normalization, we first need to understand the biases found in high-throughput sequencing data. Dillies et al. \cite{dillies_comprehensive_2013} describe two kinds of biases: \textit{intra-sample biases} which result in mis-measurement for individual genes, for example due to its length or GC content, and \textit{inter-sample biases} which result in mis-measurement for individual samples. Put symbolically, the \textit{observed abundance} $x_{ij}$ for sample $i$ and gene $j$ is produced as a function of the \textit{true abundance} $x^*_{ij}$, the gene-specific mis-measurement bias $\theta_j$, the sample-specific mis-measurement bias $C_i$, and some experimental noise $\epsilon_{ij}$:
\begin{equation}
x_{ij} = f \Big( x^*_{ij}, \theta_j, C_i, \epsilon_{ij} \Big).
\end{equation}
The bias $\theta_j$ is often safely ignored in DAA because it is assumed to have the same effect across all samples (though it can have a major impact on gross summary statistics like alpha diversity \cite{mclaren_consistent_2019}).
The bias $C_i$, called the \textit{sequencing depth}, poses a greater problem for DAA. It causes each sample vector $\textbf{x}_i$ to have an unreliable total, meaning that the observed abundances are not the true abundances.
Although technological advances (e.g., unique molecular identifiers used in single cell sequencing) can mitigate the impact of sequencing biases, their presence has a major effect on DAA: because the observed abundances are not the true abundances, absolute DA cannot be readily calculated. Thus, sequencing biases necessitate normalization.
%
%It also makes it hard to trust \textit{presential DA} because a small value of $C_i$ may imply under-sampling which could cause a gene to appear falsely absent when it is in fact present.

How do bioinformaticians account for differences in sequencing depth? In our opinion, there are, generally speaking, 3 Schools of Thought:

\begin{itemize}
\item \textbf{Absolutism:}. This approach intends to make the between-group differences for normalized abundances the same as the between-group differences for true abundances. It works by standardizing samples with an \textit{ideal reference} called an \textit{effective library size normalization factor}. Some analysts may instead use ``spike-in'' nucleotides as an experimentally-derived ideal reference \cite{deveson_representing_2016,hardwick_spliced_2016,hardwick_synthetic_2018,tkacz_absolute_2018}. Either way, absolutist methods claim to measure \textit{absolute DA}, though whether this happens will depend on the goodness of the normalization factor. An example null hypothesis would be that the normalized abundance of gene 1 (representing the true abundance) is the same for groups 1 and 2.

\item \textbf{Proportionalism:} This approach intends to make all sample vectors sum to the same total. In other words, it ensures that $\sum_j^G x_{ij} = \sum_j^G x_{hj}$ for any two samples $i$ and $h$. This can be done by dividing out the sample sums to get proportions (sometimes expressed with a per-thousand or per-million denominator, e.g., TPM), or by rarefaction \cite{mcknight_methods_2019}. Of note, proportions (or their logarithm) can also arise from a careful distributional modelling of the transcriptional activity of the cell \cite{breda_bayesian_2019}. Proportionalist methods measure \textit{proportional DA}, though this is sometimes conflated with \textit{absolute DA}. An example null hypothesis would be that the proportion of gene 1 is the same for groups 1 and 2.

\item \textbf{Compositionalism:} This approach also intends to make all samples comparable along a common scale. It works by recasting sample vectors with respect to an internal ``reference frame'' that helps interpret any differences between samples \cite{morton_establishing_2019}. The reference need not be an ideal reference. Most often, the geometric mean of the sample serves as the reference, though one could use an individual feature or feature set instead \cite{quinn_field_2019}. For simplicity's sake, we focus on the centered log-ratio (clr) transform \cite{aitchison_statistical_1986}:
\begin{equation}
    \textrm{clr}(\textbf{x}_i) = \log \Bigg( \frac{[x_{i1}, \dots, x_{iG}]}{\sqrt[G]{\Pi_{j=1}^G x_{ij}}} \Bigg).
\end{equation}
Any constant factor $C_i$ would cancel between the numerator and denominator of the log-ratio. As such, the clr transform corrects for differences in sequencing depth (although it cannot correct for zeros caused by under-sampling \cite{lovell_counts_2020}).  Interestingly, the geometric mean resembles an effective library size normalization factor, and so the clr can (confusingly) also be considered a normalization \cite{quinn_understanding_2018}. An example null hypothesis would be that the transformed abundance of gene 1 is the same for groups 1 and 2. Whether the transformed abundances represent absolute abundances depends on whether the analyst considers the CLR to be a normalization. Traditionally, compositionalists do not \cite{boogaart_fundamental_2013}.
\end{itemize}

\begin{figure}[H]
\centering
\scalebox{1}{
\includegraphics[width=(.6\textwidth)]{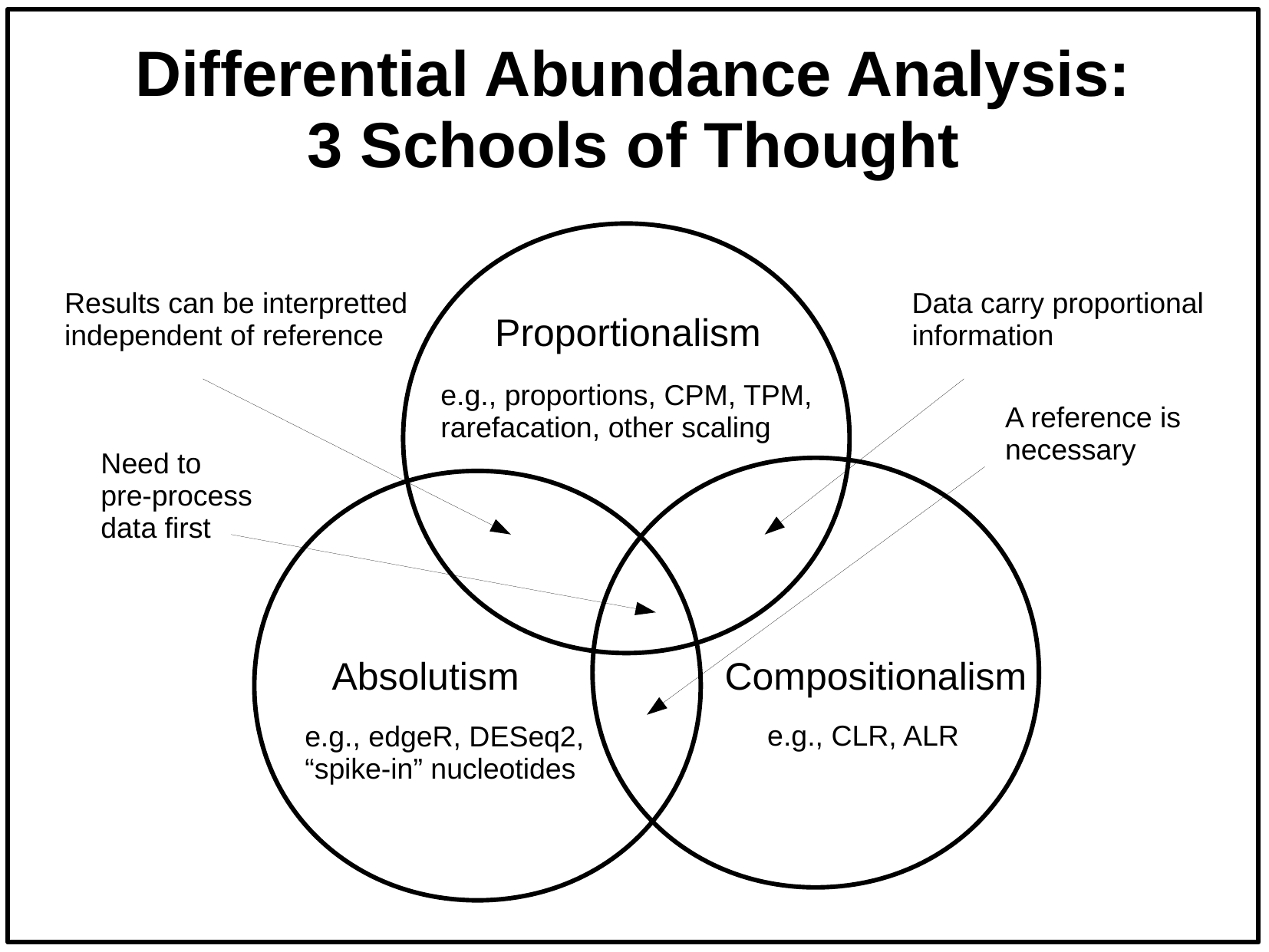}}
\caption{When it comes to differential abundance analysis, there are, generally speaking, 3 Schools of Thought. All Schools agree it is necessary to pre-process sequence count data before analysis. However, they differ in the philosophy and approach used for pre-processing. The Venn diagram identifies some beliefs held in common by the different Schools of Thought.}
\label{fig:venn}
\end{figure}

How do these Schools of Thought think about \textit{normalization}? The proportionalists consider their methods to be normalizations because they claim to divide out a source of bias $C_i$. Some proportionalist methods try to adjust for the other source of bias $\theta_j$ too, for example by dividing out fragment length (e.g., RPKM \cite{mortazavi_mapping_2008}).  The abolutists consider their methods to be normalizations because their methods are used to estimate the true between-group differences in gene abundance. Absolutists may not consider proportionalist methods to be \textit{bona fide} normalizations because an ideal reference is not used. Compositionalists may also consider a method to be a normalization if it can be used to estimate the true between-group differences in gene abundance. The clr may or may not be considered a normalization, depending on the authors using it. Quinn et al. \cite{quinn_understanding_2018} make the distinction between a \textit{clr transformation} and a \textit{clr `normalization'} which, while mathematically equivalent, differ in the motivation behind their use. Figure~\ref{fig:venn} shows a Venn diagram of some positions held by the 3 Schools of Thought.

\section{The Limitations of These Approaches}

We argue that neither the absolutist, the proportionalist, nor the compositionalist approaches are appropriate for all analyses. We have 3 reasons: (1) absolute DA requires unverifiable assumptions, (2) proportional DA is an unreliable construct, and (3) univariate models are overly reductionist. These limitations are explained below.

\subsection{Limitation 1: Absolute DA requires unverifiable assumptions}

A normalization factor removes inter-dependence by serving as an ideal reference against which to compare the observed gene counts. Imagine we knew the identity of a true ``house-keeping gene'' that was equally present in all samples. For example, let gene $c$ from the set $[a, b, c]$ be an ideal house-keeper. By definition of being a house-keeper, the absolute value of $c$ -- and not its proportional value -- is fixed. When a house-keeper is available, we can use its proportional value to make sense of the other genes in absolute terms. For example, we can analyze $[a, b]$ conditional upon $c$ \cite{aitchison_log_1984}, and if the ratio of, say, $a$-to-$c$ is bigger in one group, then $a$ is \textit{actually} more abundant. Thus, true house-keepers allow us to measure absolute DA.

When the identity of an ideal house-keeper is unknown, one could make a guess at calculating a synthetic ``house-keeper'' to serve the same purpose. Making this calculation requires making an assumption about the distribution of the data. Popular effective library normalizations like TMM \cite{robinson_scaling_2010} and DESeq2 \cite{anders_differential_2010} -- as well as clr `normalization' \cite{quinn_understanding_2018} -- assume that the majority of genes remain unchanged between samples. Under this assumption, a measure of the central tendency of the samples, such as their geometric means or medians, will act like a good house-keeper (while having an additional advantage over single house-keepers in that individual gene fluctuations are averaged out). When the assumption holds, normalization factors, like ideal house-keeping genes, allow us to measure absolute DA.

Does it make sense to assume that the majority of genes remain unchanged between samples? Maybe. It is at least a parsimonious assumption. Our point is not to advise when and where the assumption holds; perhaps this judgement is best left to the scientist performing the experiment. Rather, our point is to highlight that the assumption itself -- whether true or not -- is not verifiable without extensive experimentation that goes beyond routine sequencing (c.f., \cite{marguerat_quantitative_2012}). In order to verify whether the majority of genes are unchanged, one would need prior knowledge about the majority of genes. This would require absolute quantification of gene abundance, which, if done by sequencing, would require a way to verify whether the majority of genes are unchanged. This, in turn, would require prior knowledge about the majority of genes (and so on).

Contrary to popular belief, the clr does not solve the unverifiable assumption problem. One can think of normalization as an attempt to find a reference that answers the question, ``What is the ideal per-sample scaling factor?''. One could use compositional methods to answer this question \cite{wu_finding_2017}. When used this way, compositional methods like the clr also imply an assumption \cite{erb_how_2016}, even if it is less often acknowledged explicitly.

\subsection{Limitation 2: Proportional DA is an unreliable construct}

Gene counts obtained from sequencing are inter-dependent. A sequencer only has a limited number of reads available to count DNA molecules. When there are more DNA molecules than reads, genes compete with one another to be counted: an increase in the abundance of one gene will ``steal'' reads from the others, causing all other genes to appear less abundant.\footnote{For some low biomass samples, there may be more reads than DNA molecules, which may allow the analyst to quantify absolute gene abundances \cite{cruz_equivolumetric_2021}.}

In other words, inter-dependence makes proportional DA an unreliable construct. When we measure differences in gene proportion between groups, it tells us very little about the gene itself. For 3 genes, $[a, b, c]$, the question as to whether the fraction of $a$ differs between the groups will depend on the amount of $b$ and $c$. As such, differences in $a$ could appear and disappear just by changing the values of $b$ and $c$. In fact, $a$ could show proportional DA even when it is equally present in all samples! When looking at genes in isolation, one can easily construct such examples where the change in proportion will tell you the opposite of the change in gene transcription. Figure~\ref{fig:datypes} gives a thorough example.

It is sometimes argued that molecular proportions are all that count for the stoichiometry of the cell (or that microbial proportions are all that count for the community of a microbiome), and thus we should make proportions the basis of analysis. However, turning gene counts into proportions does \textit{not} remove the inter-dependence. In fact, as proportions, genes must again compete for representation within the fraction: having more of one gene will cause the percentage of all other genes to decrease too. This is sometimes called the unit-sum constraint because the data are constrained to sum to one unit \cite{gloor_its_2016,gloor_microbiome_2017}. Data that are subject to a unit-sum constraint are \textit{compositional} \cite{boogaart_fundamental_2013}. Yet, it has been argued that the data cannot be considered \textit{compositional} because there is always a substantial and variable proportion of reads that cannot be annotated \cite{jeganathan_statistical_2021}. However, incompletely annotated data are still \textit{compositional} because the cause of the unit-sum constraint -- i.e., the competition for representation within a sample -- remains in effect. For example, the gene set $[a, b, c]$ would still be interdependent regardless of whether a fourth gene $[d]$ was measured.
We do not think that the classical compositional treatment of rock samples is fundamentally different from the compositional modelling of sequencing experiments (a similarity that is suggested by the analogous units of parts per million vs. reads per million).\footnote{In fact, there is a simple analogy between the mass of each part in a geochemical sample and the reads of a sequencing experiment mapped to different microbial genomes. We can compare the total number of reads before mapping to the weight of the rock sample analyzed. A single read mapped to a gene or microbial genome then corresponds to the minimum amount of an oxide or other chemical compound present in the rock sample that our technology allows us to detect (e.g., if the spectrometer allows a detection of 1 part per million, this unit would get multiplied with the mass of the input material). In chemical analysis, the only substances that can be detected are those that are queried for, in the same way that only microbes can be detected whose references are available for mapping. It does not matter here that the total number of mapped reads is not predetermined for each sample (the same way that the summed weights of all substances we query for isn't a random variable of interest). Although the sequencing experiment delivers read counts whereas the compounds in a rock sample are directly assigned densities via spectroscopy, we could, in principle, recast these densities as counts of detectable units. Clearly, the genomic counts will be fewer and the number of assigned and unassigned parts greater, but the qualitative aspect of our analogy remains intact.}

Contrary to popular belief, the clr does not solve the inter-dependence problem. Although proportions transformed by the clr have real number values, they are not free to take arbitrary values: their sum must equal zero (geometrically speaking, they fall on a hyperplane in the space of the absolute data).  When we consider the clr transformation of $[a, b, c]$ -- which we denote $[a', b', c']$ -- the value of $a'$ still depends on $b$ and $c$. Like with proportions, differences in $a'$ could appear and disappear just by changing the values of $b$ and $c$.

\subsection{Limitation 3: Univariate models are overly reductionist}

Beyond the more nuanced mathematical arguments against DAA, there are some good biological reasons to seek alternatives. We consider two related reasons: DAA is overly reductionist and furthmore ignores biological stoichiometry.

First, DAA is overly reductionist. In cells, gene products (i.e., proteins) do not work in isolation. They coordinate with one another to form highly organized functional modules, for example by defining a sequence of chain reactions in a metabolic pathway \cite{ashburner_gene_2000,kanehisa_kegg_2000}. Within an ecosystem, including a microbiome sample, species are likewise linked, competing for representation within ecological niches. Microbes also cooperate with one another, and with their hosts \cite{trivedi_plantmicrobiome_2020}, to form symbiotic ecological communities, where some species ``cross-feed'' off the metabolic products of others \cite{mcdonald_application_2020}. While DAA fails to model \textit{technical interdependence}, it also fails to model this \textit{biological interdependence}.

Second, DAA ignores biological stoichiometry, which is broadly defined as the study of the balance of energy and multiple chemical elements in living systems \cite{elser_biological_2000}. In cells, RNA production capacity varies with the size of the cell, whereby two cells of the same type could have vastly different amounts of total RNA without any difference in phenotype \cite{padovan-merhar_single_2015}. Yet, a duplication of just some chromosomes tends to be more deleterious than a duplication of all chromosomes, suggesting that a change in \textit{gene balance} causes more harm than a change in total \textit{gene dosage} \cite{birchler_gene_2010,birchler_gene_2012}. In ecosystems, species within a niche have in common similar nutrient requirements, meaning that the stoichiometry of nutrients is a key aspect of their ecological niche \cite{elser_biological_2000,elser_biological_2003}. Likewise, the stoichiometry of species -- or the \textit{entropy} thereof, known in ecology as a type of \textit{diversity} \cite{hill_diversity_1973} -- can itself describe an ecosystem \cite{huttenhower_structure_2012}. For example, the total health of the gut microbiome -- itself an ecosystem -- may associate with the stoichiometry of Firmicutes species to Bacteroidetes species \cite{crovesy_profile_2020,magne_firmicutesbacteroidetes_2020}. DAA cannot model these stoichiometric complexes.

Mesoscale descriptors like modules and communities help define a unit of analysis that strikes a useful compromise between overly reductionist microscale descriptors (e.g., gene-level differential abundance) and overly broad macroscale descriptors (e.g., gene co-expression network topology) \cite{hall_co-existence_2019}. Existing methods like functional network analysis \cite{langfelder_wgcna:_2008}, multivariable ordination \cite{bodein_generic_2019}, and pathway scoring \cite{beykikhoshk_deeptriage_2020} can establish mesoscale features that describe the data more comprehensively than individual gene abundances. Meanwhile, biological stoichiometry offers another lens through which to understand living systems, including molecular and ecological interactions. Both perspectives are missing from DAA, but are present in our alternative framework.

\section{An Alternative Framework}

We agree with recent literature that recommends the use of ratios, especially sparse ratios, for the statistical analysis of gene expression and microbial abundance \cite{morton_balance_2017,silverman_phylogenetic_2017,washburne_phylogenetic_2017,rivera-pinto_balances:_2018,calle_statistical_2019,quinn_interpretable_2020}. Building on these works, we formulate an alternative framework called \textit{ratio-based biomarker analysis} that solves the 3 limitations of DAA:
\begin{enumerate}
    \item Ratio-based biomarkers are reliable constructs. Given a composition $[a, b, c]$, the value of $c$ has no influence on the ratio $\log \frac{a}{b}$. Meanwhile, the log-scale presents itself as a natural choice for proportional data. It leaves the absolute value of the ratio unchanged when exchanging the numerator with the denominator, while further reducing the observed skew in the abundance distributions. It also turns products into sums, thus allowing a simple additive model to capture multiplicative interactions among stoichiometric complexes.
    
    \item Ratio-based biomarkers do not require unverifiable assumptions. The denominator serves as a reference against which to interpret the numerator. This reference can be used in place of a normalizing reference. The reference need not be an ideal reference.
    
    \item Ratio-based biomarkers are less reductionist. Ratio-based biomarkers establish mesoscale descriptors that measure the activity of some genes (the numerator) with respect to other genes (the denominator). By definition, ratio-based biomarkers describe stoichiometry.
\end{enumerate}

%Log-contrast analyses can be aggregation-free or aggregation-based.
The simplest ratio-based biomarker is a \emph{pairwise log-ratio}, where the numerator and denominator each contain only 1 gene. Just as DAA measures the association between an outcome and a gene, \textit{differential ratio analysis} measures the association between an outcome and a gene ratio \cite{mandal_analysis_2015,walach_robust_2017,erb_differential_2018}, where a statistical test is performed for each pairwise log-ratio (with or without a subsequent summary step that distills the results into gene-level attribution scores).\footnote{Note that having $p$ genes implies $p(p-1)/2$ unique ratios and thus $O(p^2)$ statistical tests, making pairwise log-ratio analyses vulnerable to Type II errors following p-value adjustment.}
%
%A notable limitation here is that statistical power depletes quickly, since for $G$ genes there are $G*(g-1)/2$ ratios (though like DAA an empirical Bayes framework could bring more power to the analysis \cite{ionas}). %Alternatively, one could use prior knowledge to focus only on ratios containing genes-of-interest, for example genes belonging to a certain molecular pathway (cite field guide).
%

\begin{figure}[H]
\centering
\scalebox{1}{
\includegraphics[width=(.6\textwidth)]{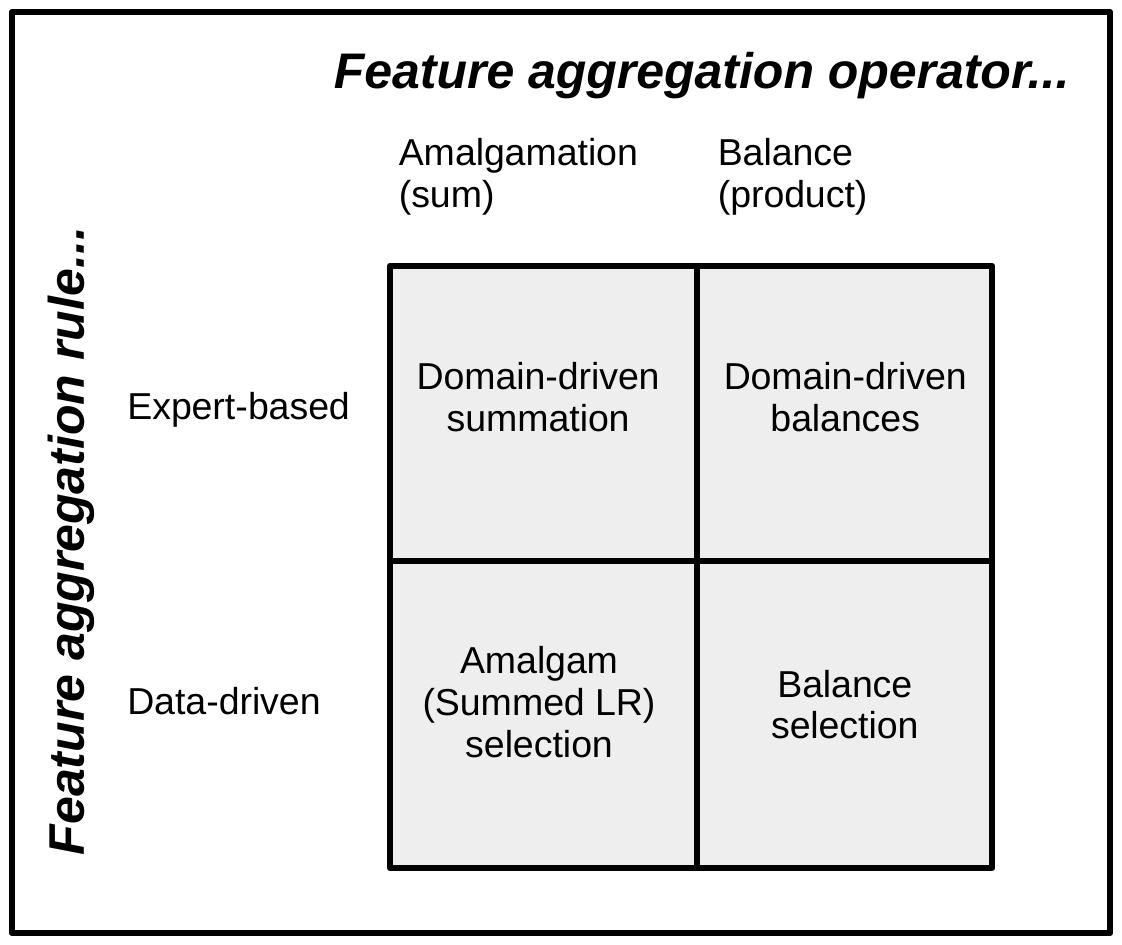}}
\caption{It is helpful to classify aggregation methods along two descriptive axes. The first describes the aggregation rule: expert-based vs. data-driven. The second describes the operation: addition vs. multiplication. Ratio-based biomarkers based on addition are called summed log-ratios, while ratio-based biomarkers based on multiplication are called balances.}
\label{fig:aggregate}
\end{figure}

It is sometimes desirable to produce multi-part ratio biomarkers, containing 2 or more genes in the numerator or denominator. %The gene abundances may be combined (i.e., \textit{aggregated}) using the addition or multiplication operation.
%When we say \textit{aggregation}, . 
Several methods exist for \textit{aggregating} genes, by which we mean \textit{combining input features to form new composite features}. Figure~\ref{fig:aggregate} arranges aggregation methods along two descriptive axes. The first axis describes the aggregation rule: Are features aggregated based on prior expert knowledge (e.g., by grouping bacteria by phylogeny)? Or are they aggregated using an algorithm that finds patterns in the data (e.g., by machine learning)? The second axis describes 
the operation used to aggregate the parts: Are features aggregated by addition (e.g., arithmetic mean)? Or are they aggregated by multiplication (e.g., geometric mean)?

Ratio-based biomarker analysis is algebraically straightforward, and several open-source software tools already exist to implement this analysis. Below, we survey existing methods, and discuss how the flexibility of ratio biomarkers allows them to be used many other bespoke analyses, including dimension reduction and multi-omics data integration.

\subsection{Aggregation by product}

When features are aggregated by multiplication, the resulting ratio-based biomarker corresponds to a log-linear combination of variables (with coefficients summing to zero to ensure scale invariance). We highlight an important special case: when the numerator and denominator of our ratio correspond to simple geometric means of one or more features. Such ratios are
%A ratio-based biomarker between geometric means is 
called \textit{balances} \cite{egozcue_groups_2005}, and can be written as:\footnote{Note that the original definition involves a prefactor as balances are related to orthonormal bases of the simplex. The prefactor is practically unnecessary for balance selection, as it would get absorbed into the regression coefficient.}
\begin{align} \label{eq:balance}
z_{i} &= \log \left( \frac{(\prod_{j \in J^+} x_{ij})^{\frac{1}{G^+}}} {(\prod_{j \in J^-} x_{ij})^{\frac{1}{G^-}}} \right)
\\ \nonumber %\label{eq:balance2}
&= \frac{1}{G^+}\sum_{j \in J^+} \log x_{ij} - \frac{1}{G^-} \sum_{j \in J^-} \log x_{ij},
\end{align}
where $J^+$ and $J^-$ denote two disjoint sets of genes, of sizes $G^+$ and $G^-$, respectively. Although the equation may look daunting, it just describes a log-ratio between two geometric means, where each average is the average of one gene set. There is one balance score $z_i$ for each sample $i$, and together they form a new feature vector $\textbf{z}$. Balances can be constructed according to a mathematical rule \cite{egozcue_isometric_2003,egozcue_groups_2005} that one can derive from prior knowledge \cite{silverman_phylogenetic_2017,washburne_phylogenetic_2017} or from the data itself \cite{pawlowsky-glahn_principal_2011,morton_balance_2017,quinn_interpretable_2020}.

Having fewer genes within each set would constitute a sparser balance, reducing the cognitive load required for interpretation. Sparsity is also advantageous from a statistical point of view, typically helping to prevent model overfit. The special case where $G^+ = G^- = 1$ reduces to the pairwise log-ratio between two individual genes. Balances therefore define a strictly greater class of features, offering the potential for much more flexible families of models.\footnote{Given $G$ input genes, one can derive $\sim G^2$ pairwise log-ratios, but $\sim (2^G)^2$ balances, a vastly greater number.}
Note that any scaling factor applied to $\textbf{x}_i$ cancels out in the ratio of Eq. \ref{eq:balance}, and so balances are the same for both relative and absolute data.

Recently, interest has gathered around \textit{balance selection}, which involves identifying maximally predictive balances algorithmically \cite{rivera-pinto_balances:_2018,quinn_interpretable_2020,susin_variable_2020,gordon-rodriguez_learning_2021}.
The \textsf{selbal} algorithm uses forward stepwise linear regression to identify a single balance associated with an experimental condition \cite{rivera-pinto_balances:_2018}. We can express the final model by analogy to the generalized linear model used in DAA
\begin{equation}
    \textbf{y} \approx \phi \Big( \beta \textbf{z}  + \beta_0 \Big).
\end{equation}
The balance feature $\textbf{z}$ is \textit{predictive} if its effect size $\beta$ exceeds some threshold. In practice, the strategy for \textit{validating} the association follows from the algorithmic modelling culture, where test set verification is used to measure out-of-sample model performance empirically \cite{breiman_statistical_2001} (i.e., instead of reporting a p-value). The \textsf{selbal} package makes balances easy to learn on low-to-moderate dimensional data ($G < 100$ genes).

\begin{lstlisting}
devtools::install_github(repo = "UVic-omics/selbal")
library(selbal)
model <- selbal(train.x, train.y)
yhat <- predict(model, test.x)
\end{lstlisting}

As for larger, high-dimensional data sets ($100 < G < 100,000$ genes), a recent method called \textsf{codacore} achieves much the same goal more efficiently \cite{gordon-rodriguez_learning_2021}. This method uses back-propagation with a network-inspired architecture instead of stepwise regression, resulting in fast runtimes on very high-dimensional data sets. In addition, \textsf{codacore} features a single regularization term, controlled by the argument \textsf{lambda}, which can be used to control the sparsity of the learned balances.

\begin{lstlisting}
devtools::install_github(repo = "egr95/R-codacore", ref="main")
library(codacore)
model <- codacore(train.x, train.y, lambda = 1)
yhat <- predict(model, test.x)
\end{lstlisting}

Like with selbal, any learned balance should be validated using test set verification. If the balance achieves good predictive accuracy in an external test set, then it can be considered a \textit{bona fide} ratio-based biomarker, and interpreted in place of univariate DAA results.

\subsection{Aggregation by sum}

The summing of parts is called an \emph{amalgamation}. A ratio-based biomarker using amalgamations is called a \textit{summed log-ratio} (SLR) \cite{greenacre_amalgamations_2020,greenacre_comparison_2021}. An SLR has the form
\begin{align} \label{eq:slr}
z_{i} = \log \left( \frac{\sum_{j \in J^+} x_{ij} } {\sum_{j \in J^-} x_{ij}} \right),
\end{align}
where again, $J^+$ and $J^-$ denote two disjoint sets of genes. As above, having fewer genes in each set results in a sparser ratio that is arguably easier to interpret. Like balances, there is one SLR score for each sample, and together they form a new feature vector $\textbf{z}$.

SLRs were proposed to solve two perceived limitations with balances \cite{greenacre_isometric_2019}. First, a log-ratio of geometric means -- upon which balances depend -- is undefined if any element equals zero. Such zeros are handled naturally by the sum. 
Second, a log-ratio of geometric means is not a ``balance'' in the plain English sense of the word and small values have a strong influence on the outcome. A \textit{balance}, as in a \textit{scale}, compares the \textit{additive} -- not \textit{multiplicative} -- totals of two sets.

Although the SLR was conceived separately from the genomics literature, it is already familiar to our field. For example, the Firmicutes-to-Bacteroidetes ratio, studied as a biomarker for obesity \cite{crovesy_profile_2020,magne_firmicutesbacteroidetes_2020}, is a ratio between disjoint sets of bacteria taxa aggregated by sum based on their respective phyla membership. The log of this ratio is an SLR, whose association with an outcome could be expressed by analogy to the generalized linear model used in DAA.
\begin{equation}
    \textbf{y} \approx \phi \Big( \beta \textbf{z}  + \beta_0 \Big).
\end{equation}
An SLR feature $\textbf{z}$ is \textit{predictive} when its effect size $\beta$ exceeds some threshold, and is validated empirically using out-of-sample model performance.

In the case of the Firmicutes-to-Bacteroidetes SLR, the aggregation rule is defined by prior expert knowledge about microbe phylogeny. Like balances, it is possible to establish data-driven SLRs too, with recent studies proposing new methods for SLR selection. The \textsf{amalgam} package uses evolutionary algorithms to identify an SLR that optimizes a regression fit \cite{quinn_amalgams_2020}.

\begin{lstlisting}
devtools::install_github(repo = "tpq/amalgam")
library(amalgam)
model <- amalgam(train.x, train.y, asSLR = TRUE, n.amalgam = 2)
yhat <- predict(model, test.x)
\end{lstlisting}

\begin{figure}[H]
\centering
\scalebox{1}{
\includegraphics[width=(.6\textwidth)]{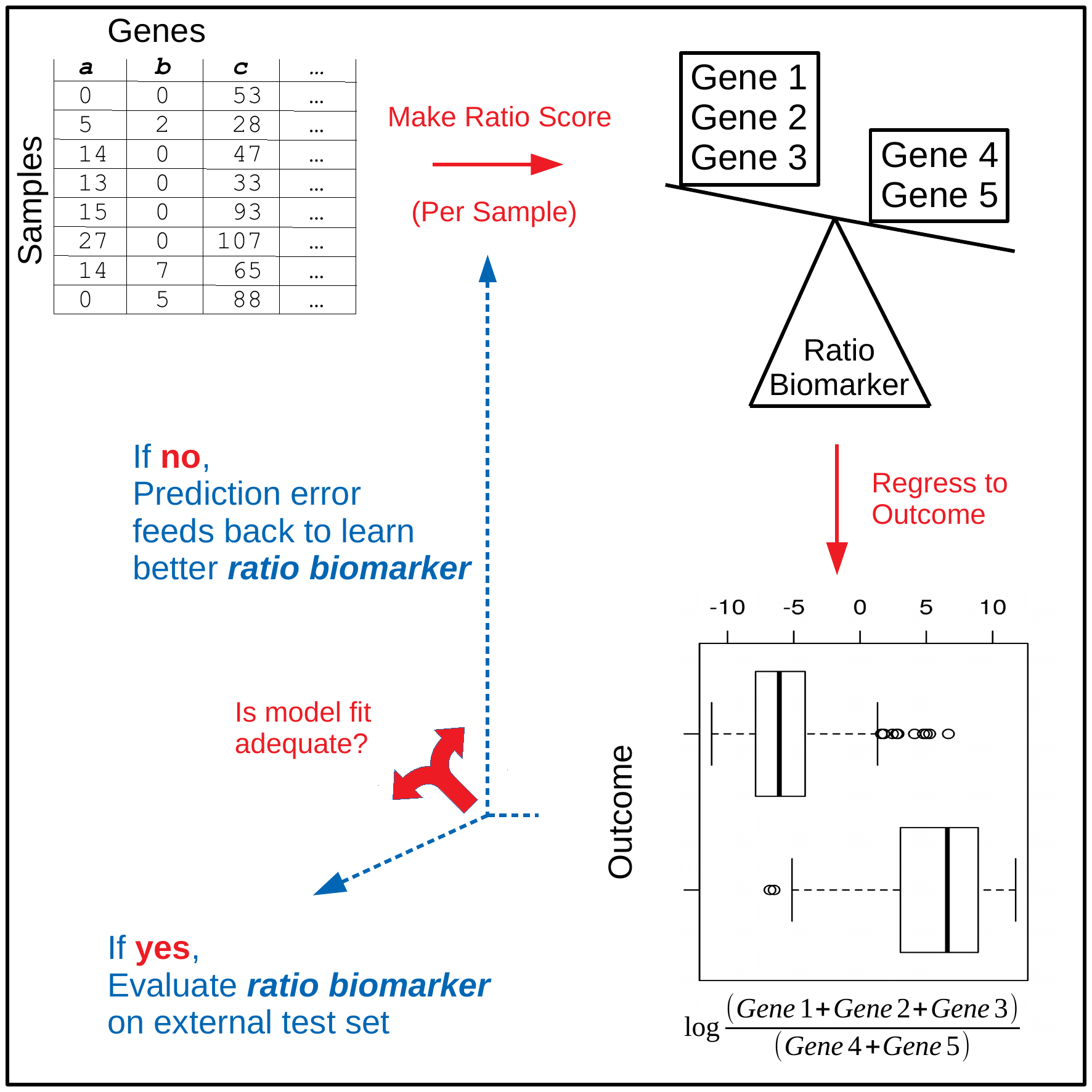}}
\caption{A high-level overview of how algorithms can learn predictive ratio-based biomarkers. First, the input data are used to randomly initialize a ratio-based biomarker.Next, the biomarker score is regressed against a continuous or categorical outcome, and a loss function is used to summarize the goodness of the fit. Finally, the algorithm constructs a new ratio-based biomarker and the cycle repeats. Once the fit is adequate, the algorithm stops and the best biomarker is returned. Different optimization algorithms can be used to find the best biomarker.}
\label{fig:DAlimits}
\end{figure} 

The \textsf{codacore} package can also learn SLRs by setting the argument \textsf{logRatioType="SLR"} \cite{gordon-rodriguez_learning_2021}. As before, this method enjoys fast runtimes on very high-dimensional data sets, and tends to find very sparse biomarkers.

% devtools::install_github(repo = "egr95/R-codacore", ref="main")
% library(codacore)
\begin{lstlisting}
model <- codacore(train.x, train.y, logRatioType = "SLR", lambda = 1)
yhat <- predict(model, test.x)
\end{lstlisting}

We refer our readers to \cite{gordon-rodriguez_learning_2021} for a systematic benchmark of balance selection, SLR selection, and other state-of-the-art methods.

\section{The Flexibility of Ratio Biomarkers}

Ratio-based biomarkers (RBB) are highly flexible. Beyond replacing DAA, they can be further used for many other bespoke analyses. In this section, we show how RBB can add value to dimension reduction and multi-omics data integration alike, bringing greater interpretability to multivariable models. Our goal here is not to introduce new methodology; rather, we aim to unify conventional multivariable analysis with our proposed RBB framework.

It is increasingly common to collect multiple ``-omics'' data sets, %. When different quantities for the same features are obtained, this is known as multi-mode data. Alternatively, one may sample different features in parallel, 
e.g., microbe abundances and metabolite abundances \cite{smolinska_volatile_2018,tang_multi-omic_2019,lloyd-price_multi-omics_2019,yachida_metagenomic_2019,franzosa_gut_2019}. In addition to analyzing each of the data sets separately, one could perform a single analysis of the combined data, revealing associations within and between the different modalities \cite{le_cao_sparse_2008,le_cao_sparse_2011,meng_dimension_2016,paliy_application_2016}. In multi-omics data integration, the task is complicated because both data sets contain their own intra-sample and inter-sample biases \cite{morton_learning_2019}. Intuitively, each data set should get normalized separately before an analysis \cite{le_deep_2020,quinn_examining_2021}, for example by using the clr. However, if we accept that a single normalization is problematic, then we might agree that the use of two normalizations is doubly problematic \cite{quinn_examining_2021}.
%
%Given the microbe abundances $\textbf{T}$ and metabolite abundances $\textbf{U}$, one could learn a function $f$ that relates their normalized forms
%
%\begin{equation}
%    \textrm{clr}(\textbf{U}) = f(\textrm{clr}(\textbf{T}))
%\end{equation}

As a flexible framework, RBB can be leveraged to learn multivariable biomarkers that do not depend on normalization factors. For example, given the microbe abundances $\textbf{T}$ and metabolite abundances $\textbf{U}$, we could learn a microbe ratio $\textbf{z}_t$ and a metabolite ratio $\textbf{z}_u$ that are highly correlated. Together, they would form a pair of ratio-based biomarkers that describe the commonality between two modes of data, e.g., a bacteria stoichiometry that is maximally predictive of a metabolite stoichiometry. The interpretation of these $\mathbf{z}$ would not depend on any normalization factors by definition of them being scale-invariant log-ratios.

We extend RBB to multivariable applications by learning sparse and interpretable ratio biomarkers that serve as drop-in replacements for other low-dimensional latent representations. Put symbolically, we simply regress:
\begin{equation}
    \textbf{h} \approx \beta \textbf{z} + \beta_0
\end{equation}
where the RBB \textbf{z} is a replacement for the latent component \textbf{h}. %Although $\mathbf{h}$ may suffer from normalization biases, . 
This approach can be used for dimension reduction (e.g., when \textbf{T} is regressed onto a function of \textbf{T}, or \textbf{U} onto \textbf{U}), or for multi-omics data integration (e.g., when \textbf{T} is regressed onto a function of \textbf{U}, or \textbf{U} onto \textbf{T}).

Table~\ref{tab:R2} compares the performance of several multivariable RBB algorithms, based on either (a) principal components analysis (PCA), (b) partial least squares (PLS) regression, or (c) neural network (NN) encoder-decoders.%
\footnote{For this benchmark, we use paired microbiome and metabolome abundance data, made publicly available by \cite{franzosa_gut_2019}. The data are lightly pre-processed by (a) removing any feature with $>50\%$ zero values and (b) replacing the remaining zeros with one half the detection limit. See \cite{lubbe_comparison_2021} for a comparison of zero replacement strategies.}%
\footnote{Our encoder-decoder network architecture consists of an encoder mapping the input to a 1-dimensional latent space, and a decoder mapping the latent variable to the output. For simplicity, our encoder and decoder each have a single hidden layer with 32 units and ReLU nonlinearities.}
With each algorithm, a (non-sparse) latent representation of the data is learned using off-the-shelf libraries. This representation is then approximated with a sparse RBB, using the \textsf{codacore} package. Finally, the original latent representations as well as the RBB approximations are used to reconstruct the original data, and the proportion of variance explained ($R^2$) is measured. We find that the RBB approximations are similarly powerful to the original representations, while enjoying robustness against normalization biases as well as a high level of sparsity, thus dramatically enhancing model interpretability. Our Appendix presents an in-depth and reproducible case study of the PLS example, combining the \textsf{codacore} \cite{gordon-rodriguez_learning_2021} and \textsf{mixOmics} \cite{rohart_mixomics_2017} packages into a single analysis.

\begin{table}[H]
\label{tab:R2}
\begin{center}
\begin{tabular}{cccccc}
\toprule
\shortstack{Objective \\ $\textrm{}$} & \shortstack{1-D latent \\ representation} & %\shortstack{ Original \\ \# vars} & 
\shortstack{Original \\ $R^2$} &
\shortstack{RBB \\ source} & \shortstack{RBB \\ \# vars} &  \shortstack{RBB \\ $R^2$} \\ 
\midrule
\shortstack{Dimension}    & $\mathbf{h}=\mathrm{PCA}_1(\mathrm{clr}(\mathbf{T}))$ & %53 & 
0.31 & $\mathbf{T}$ & 18 / 58 & 0.31 \\
%\midrule
\shortstack{reduction}    & $\mathbf{h}=\mathrm{PCA}_1(\mathrm{clr}(\mathbf{U}))$ & %3218 & 
0.34 & $\mathbf{U}$ & 5 / 7156 & 0.33\\
\midrule
\shortstack{Multi-omics} & $\mathbf{h}=\mathrm{PCA}_1(\mathrm{clr}(\mathbf{U}))$ & %3218 & 
0.34 & $\mathbf{T}$ & 4 / 58 & 0.31\\
%\midrule
\shortstack{integration} & $\mathbf{h}=\mathrm{PCA}_1(\mathrm{clr}(\mathbf{T}))$ & %53 & 
0.31 & $\mathbf{U}$ & 3 / 7156 & 0.27\\
\midrule
\shortstack{Dimension} & $\mathbf{h}=\mathrm{PLS}^{(\textbf{T})}_1(\mathrm{clr}(\mathbf{T}), \mathrm{clr}(\mathbf{U}))$ & %55 & 
0.30 & $\mathbf{T}$ & 18 / 58 & 0.30 \\
\shortstack{reduction} & $\mathbf{h}=\mathrm{PLS}^{(\textbf{U})}_1(\mathrm{clr}(\mathbf{T}), \mathrm{clr}(\mathbf{U}))$ & %3351 & 
0.35 & $\mathbf{U}$ & 8 / 7156 & 0.34\\
\midrule
\shortstack{Multi-omics} & $\mathbf{h}=\mathrm{PLS}^{(\textbf{U})}_1(\mathrm{clr}(\mathbf{T}), \mathrm{clr}(\mathbf{U}))$ & %3351 &
0.35 & $\mathbf{T}$ & 6 / 58 & 0.31 \\
\shortstack{integration} & $\mathbf{h}=\mathrm{PLS}^{(\textbf{T})}_1(\mathrm{clr}(\mathbf{T}), \mathrm{clr}(\mathbf{U}))$ & %55 & 
0.26 & $\mathbf{U}$ & 19 / 7156 & 0.31 \\

\midrule
\shortstack{Dimension} & $\mathbf{h}$, where $\mathbf{T} \overset{\scriptscriptstyle{NN}}{\rightarrow} \mathbf{h} \overset{\scriptscriptstyle NN}{\rightarrow} \mathbf{T}$ & %58 & 
0.35 & $\mathbf{T}$ & 20 / 58 & 0.31\\
%\midrule
\shortstack{reduction} & $\mathbf{h}$, where $\mathbf{U} \overset{\scriptscriptstyle{NN}}{\rightarrow} \mathbf{h} \overset{\scriptscriptstyle NN}{\rightarrow} \mathbf{U}$ & %7156 & 
0.40 & $\mathbf{U}$ & 7 / 7156 & 0.35 \\
%\midrule
\midrule
\shortstack{Multi-omics} & $\mathbf{h}$, where $\mathbf{T} \overset{\scriptscriptstyle{NN}}{\rightarrow} \mathbf{h} \overset{\scriptscriptstyle NN}{\rightarrow} \mathbf{U}$ & %58 & 
0.39 & $\mathbf{T}$ & 7 / 58 & 0.30\\
\shortstack{integration}                         & $\mathbf{h}$, where $\mathbf{U} \overset{\scriptscriptstyle{NN}}{\rightarrow} \mathbf{h} \overset{\scriptscriptstyle NN}{\rightarrow} \mathbf{T}$ & %7156 & 
0.34 & $\mathbf{U}$ & 3 / 7156 & 0.28 \\
%\midrule
%\shortstack{Multi-omics} & $\mathbf{U} \overset{\scriptscriptstyle{NN}}{\rightarrow} \mathbf{h} \overset{\scriptscriptstyle NN}{\rightarrow} \mathbf{T}$ & 7156 & 0.17 & $\mathbf{T}$ & 19 & 0.16\\
%\midrule

%\midrule
%\shortstack{integration}                         & $\mathbf{T} \overset{\scriptscriptstyle{NN}}{\rightarrow} \mathbf{h} \overset{\scriptscriptstyle NN}{\rightarrow} \mathbf{U}$ & 58 & 0.28 & $\mathbf{U}$ & 3 & 0.24\\
\bottomrule
\end{tabular}
\caption{Proportion of variance explained ($R^2$) for the top latent component from PCA, PLS, and NN encoder-decoders. For each latent representation, we show the $R^2$ obtained by the original model alongside a single RBB optimized to approximate that representation. Generally speaking, the $R^2$ values largely agree. Yet, based on the number of active variables, the RBB can achieve 65\%-99\% greater sparsity than a completely dense model. %We highlight the sparsity achieved by RBB; compare the number of active input variables in the original latent representations versus their RBB counterparts 
%(for PCA, PLS and NN, we consider an input variable to be active when its coefficient is greater in magnitude than $0.01$). 
While we do consider several multivariable applications of RBB, the table is by no means exhaustive.}
\end{center}
\end{table}

%These examples speak to the flexibility of the RBB framework: a single research objective can elicit multiple valid solutions. With each solution, RBB produces sparse and interpretable biomarkers that can help answer questions that DAA cannot. We believe that the myriad of ways in which RBB can be used to analyze genes, the microbiome, their metabolites, and other ``-omes'' presents our field with many exciting research opportunities.

\section{Summary}

A critical examination of differential abundance analysis (DAA) reveals intrinsic limitations that researchers should seriously consider prior to using DAA. We advise that when using DAA, researchers should explicitly acknowledge the implicit assumptions they make, and recognize that the constructs used for hypothesis testing may be confounded by inter-dependence. Alternatively, they may wish to use ratio-based biomarkers instead. These ratios provide an intuitively simple alternative that does not suffer from the limitations of DAA. Ratio-based biomarkers require no normalization and cannot be confounded by interdependence. Advantageously, the framework is highly flexible and can be readily used for bespoke analyses to  discover sparse, interpretable stoichiometric complexes that (co-)vary across time or space or experiment. The myriad of ways in which RBB can be used to analyze genes, the microbiome, their metabolites, and other ``-omes'' presents our field with many exciting research opportunities.

\section{Availability of data and code}

The data and benchmark analysis are available from \url{https://zenodo.org/record/4692004}.

\section{Acknowledgements}

TPQ thanks Thomaz Bastiaanssen for helpful discussions on theoretical ecology.

\bibliographystyle{unsrt}
\bibliography{references,references_Ionas}

\end{document}